\documentclass[12pt,a4paper,reqno%,draft
]{amsart}

\usepackage{graphicx} % Required for including images
\usepackage{amsmath,amssymb,amsthm,array}
\usepackage{amsfonts, amsmath, amsthm, amssymb} % For math fonts, symbols and

\newcommand{\LambertW}{\mathop{\mathrm{LambertW}}\nolimits}

\begin{document}
\title[ Periodic boundary, cylindrical and spherical KdV-Burgers] {On periodic boundary solutions for cylindrical and spherical KdV-Burgers equations}

\author{Alexey Samokhin}\vspace{6pt}

\address{Institute of Control Sciences of Russian Academy of Sciences
65 Profsoyuznaya street, Moscow 117997, Russia}\vspace{6pt}

\email{ samohinalexey@gmail.com}\vspace{6pt}

\begin{abstract}

For the KdV-Burgers equations for cylindrical and spherical waves  the development of a regular profile starting from an equilibrium under a periodic perturbation at the boundary is studied.
The equation describes a medium which is both dissipative and dispersive. For an appropriate combination of dispersion and dissipation  the asymptotic profile looks like a periodical chain of shock fronts with a decreasing amplitude (sawtooth waves). The development of such a profile is preceded by a head shock of a constant height and equal velocity which depends on spatial dimension as well as on integral characteristics of boundary condition; an explicit asymptotic for this head shock is found.

 \vspace{1mm}

\noindent\textbf{Keywords:} Korteweg-de Vries-Burgers equation, cylindrical and spherical waves, saw-tooth solutions,  periodic boundary conditions, head shock wave.

\noindent\textbf{MSC[2010]:} 35Q53, 35B36.
\end{abstract}

\maketitle

\section{Introduction}

 The behavior of solutions of the KdV and KdV - Burgers equations is well studied, yet they remain a subject of various recent research, \cite{key-3}--\cite{key-5} where these equations describe flat waves in one spatial dimension. But cylindrical and spherical waves also have a variety applications (eg, waves generated by a downhole vibrator). The paper is a continuation of the previous research of the author, \cite{key-6} -- \cite{key-10}.

 The well known KdV-Burgers equation for flat waves is of the form
 \begin{equation}\label{01}
     u_t=-2uu_x+\varepsilon^2 u_{xx}+\delta u_{xxx}.
    \end{equation}
 Its cylindrical and spherical analogues are
 \begin{equation}\label{02}
     u_t+\frac{1}{2t}u=-2uu_x+\varepsilon^2 u_{xx}+\delta u_{xxx}.
    \end{equation}
    and
    \begin{equation}\label{03}
     u_t+\frac{1}{t}u=-2uu_x+\varepsilon^2 u_{xx}+\delta u_{xxx}.
    \end{equation}
    correspondingly, \cite{key-1} -- \cite{key-2}.

  We consider the initial value - boundary problem (IVBP) for the KdV-Burgers equation on a finite interval:
 \begin{equation}\label{07}
 u(x,0) =f(x), \; u(a,t) = l(t),\; u(b,t) =L(t), \; u_x(b,t) =R(t), \; x\in[a,b].
\end{equation}
In the case $\delta=0$ (that is, for Burgers equation), it comes to

\begin{equation}\label{08}
u(x,0) =f(x), \; u(a,t) = l(t),\; u(b,t) =R(t), \; x\in[a,b].
\end{equation}

The case of the boundary conditions $u(a,t) = A\sin(\omega t),\; u(b,t) =0$ and the related asymptotics are of a special interest here. For numerical modelling we use $ x\in[0,b]$ for appropriately large $b$ instead of $\mathbb{R}^+$.

\section{Flat case: travelling waves}

For $t\gg 1$ equations \eqref{02} and \eqref{03} tend to \eqref{01} as well as their solutions. In particular,
the explicit form of traveling wave solutions for the flat KdV-Burgers (\ref{01}) is as follows

  \begin{equation}\label{uTWS}
  u_{\mathrm{tws}}(x,t)=\frac{3\varepsilon^4\tanh^2(\frac{\varepsilon^2(x-Vt-s)}{10\delta})}{50\delta} -
\frac{3\varepsilon^4\tanh(\frac{\varepsilon^2(x-Vt-s)}{10\delta})}{25\delta}+\frac{V}{2}-\frac{3\varepsilon^4}{50\delta}
\end{equation}

Our IVBP requires $u|_{x=+\infty}=0$; so the sole such travelling wave has a velocity $V=\frac{6\varepsilon^
4}{25\delta}.$

Note that the height of the wave \eqref{uTWS},
 $u|_{x=-\infty}-u|_{x=+\infty}=H-h=6\varepsilon^4/25\delta$ does not depend on its velocity and is completely defined by the ratio $\varepsilon^4/\delta$ which depends on the coefficients $\varepsilon,\;\lambda$ related to dispersion and dissipation.

 Also note that the equations \eqref{01}--\eqref{03} may be readily put in the form
 $w_t+\frac{n}{2t}w=\gamma w_{xx}-2ww_x+w_{xxx}$ by the change of variables $t\rightarrow t\sqrt{\delta}$, $x\rightarrow x\sqrt{\delta}$,
 $u\rightarrow -\frac{u}{2}$. Here $\gamma=\frac{\varepsilon^2}{\sqrt{\delta}}$ is the important parameter that defines a character of solutions; $n=0,1/2,1$ for flat, cylindrical and spherical waves correspondingly.

In the case $\delta=0$, the Burgers equation also has a variety of travelling wave solutions, vanishing at $x\rightarrow +\infty$. They are given by the formula

\begin{equation}\label{BTWS}
u_{\mathrm{Btws}}(x,t)=\frac{V}{2}\left[1-\tanh\left(\frac{V}{2\varepsilon^2}(x-Vt+s)\right)\right]
\end{equation}

We demonstrate that  in the case of the above IVBP the perturbation of the equilibrium state \eqref{2}, \eqref{3} ultimately becomes very similar to the form of this shock.

\section{Typical examples}

\subsection{Burgers.}

\begin{figure}[h]
 \begin{minipage}{14pc}
\includegraphics[width=14pc]{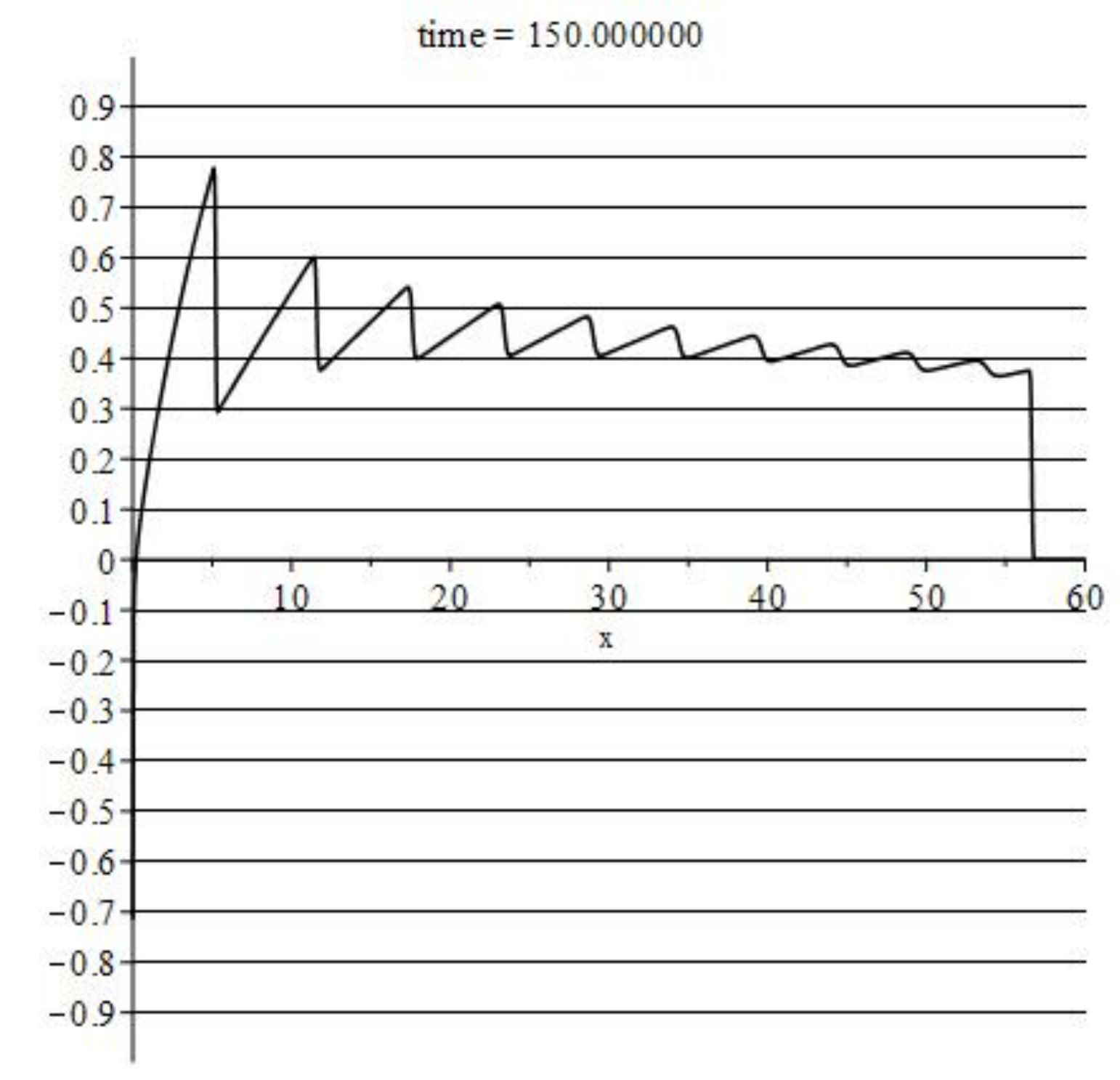}
\end{minipage}
\begin{minipage}{14pc}
\includegraphics[width=14pc]{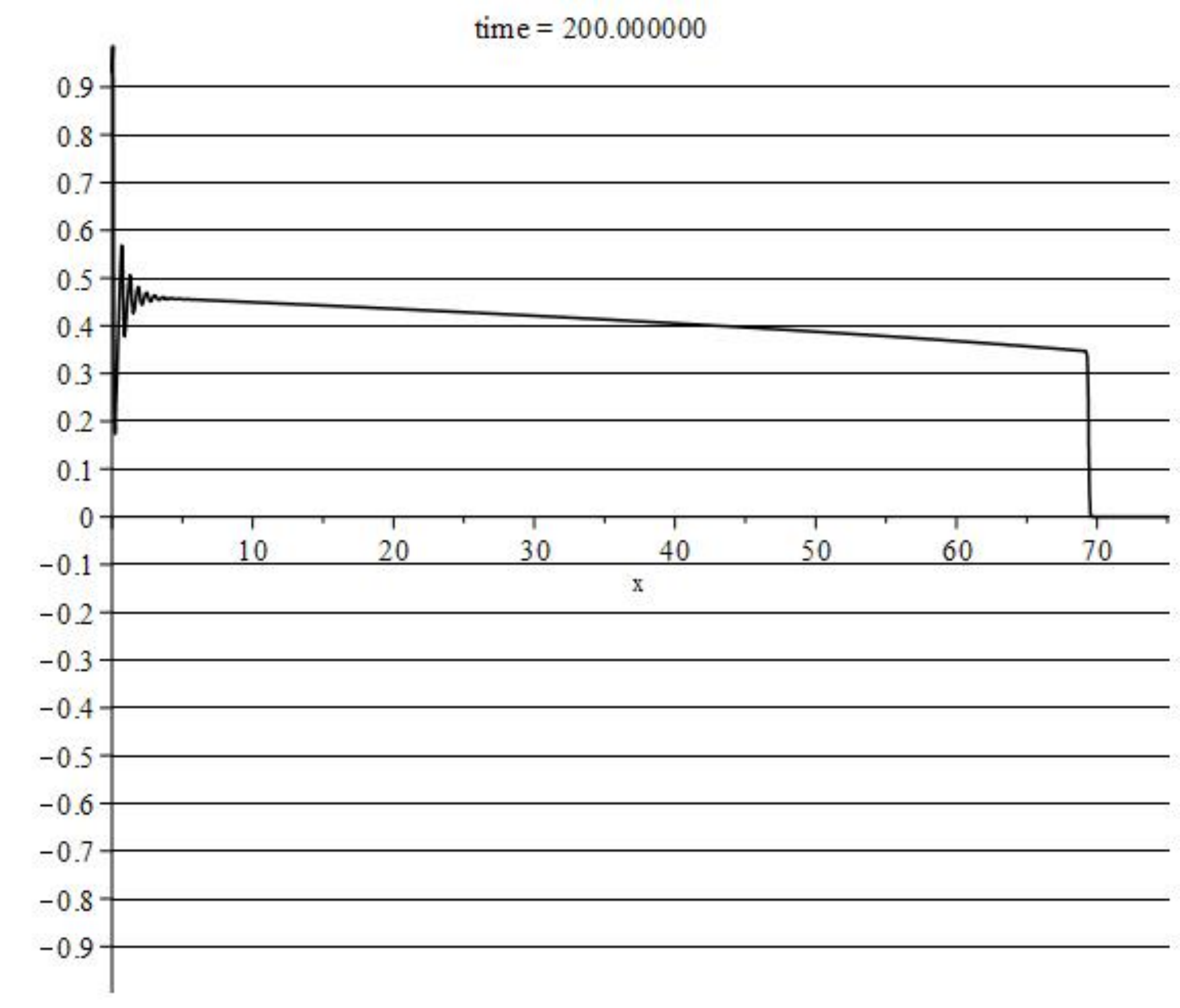}
\end{minipage}
\caption{\textsl{Cylindrical Burgers, $\varepsilon=0.1,$ \textbf{Left}: $u_0=\sin t, t=150.$ %\protect\newline
\textbf{Right}: $u_0=\sin 10t, t=200.$}}
\label{B1}
\end{figure}

\begin{figure}[h]
 \begin{minipage}{14pc}
\includegraphics[width=14pc]{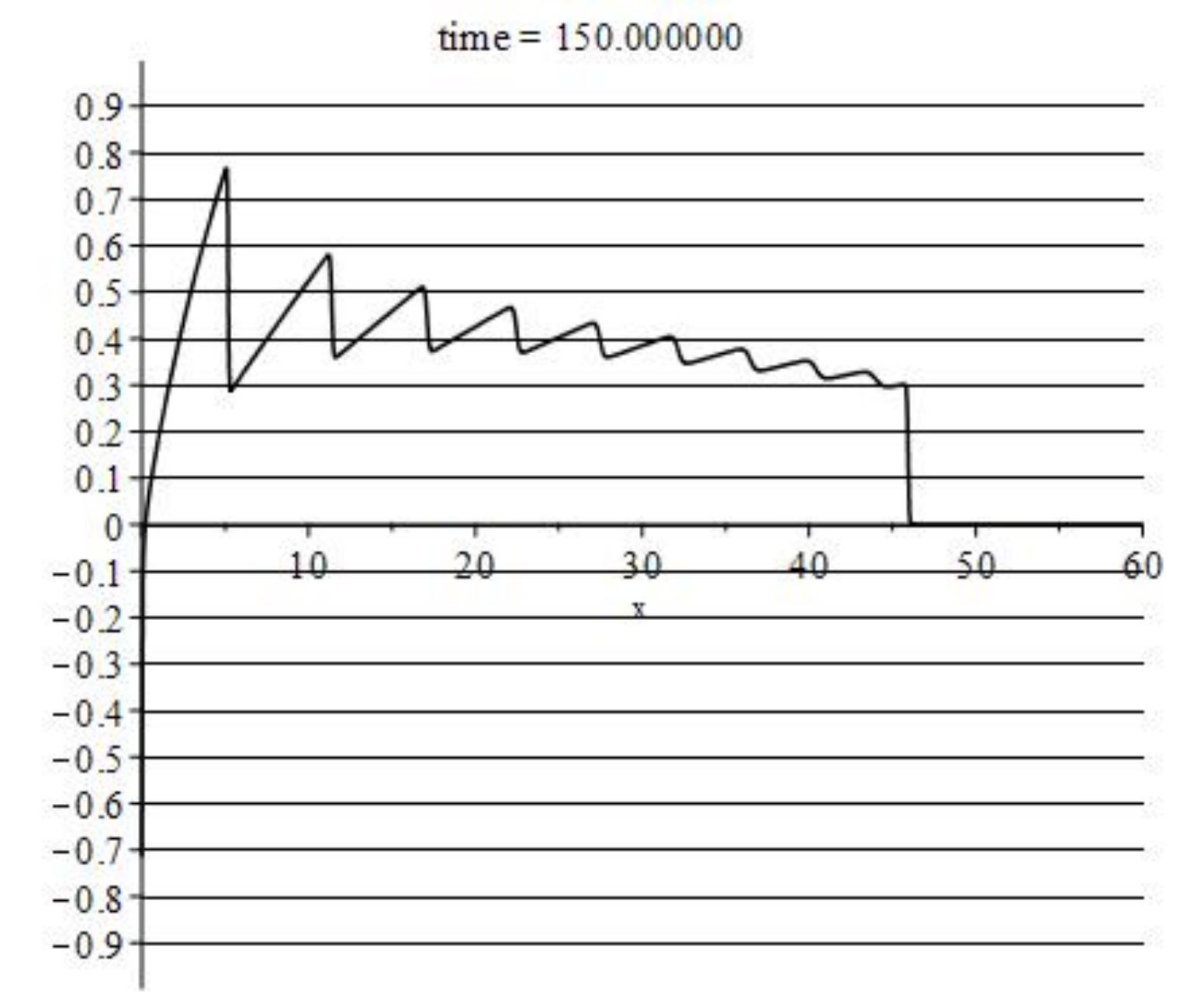}
\end{minipage}
\begin{minipage}{14pc}
\includegraphics[width=14pc]{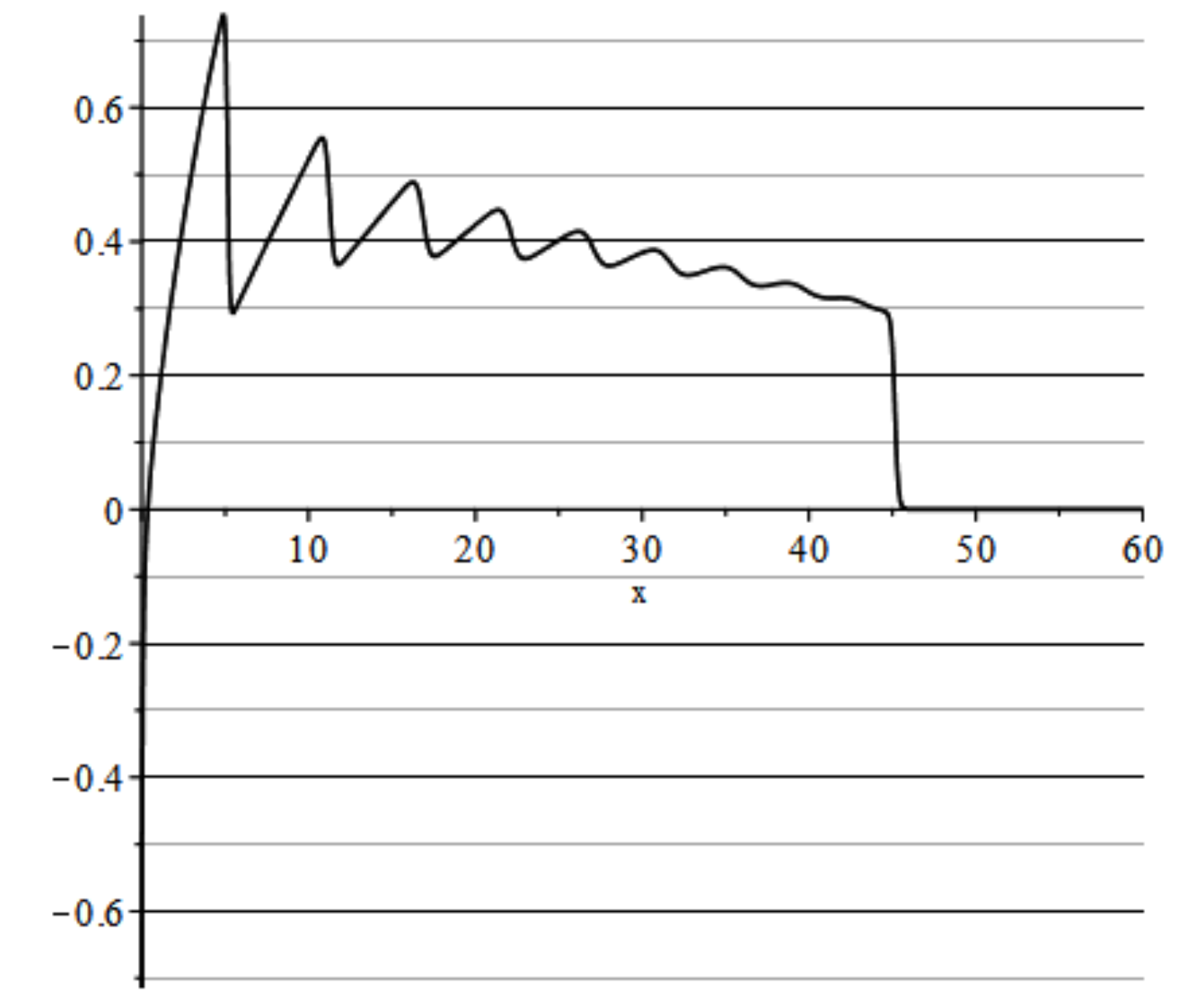}
\end{minipage}
\caption{\textsl{Spherical Burgers, $u_0=\sin t$, \textbf{Left}: $\varepsilon=0.1, t=150$  %\protect\newline
\textbf{Right}: $\varepsilon^2=0.3, t=150$ }}
\label{B2}
\end{figure}

 Here we demonstrate typical graphs for cylindrical and spherical Burgers waves, figure \ref{B1}, \ref{B2}.

\subsection{KdV-Burgers.}

\begin{figure}[b]
 \begin{minipage}{14pc}
\includegraphics[width=14pc]{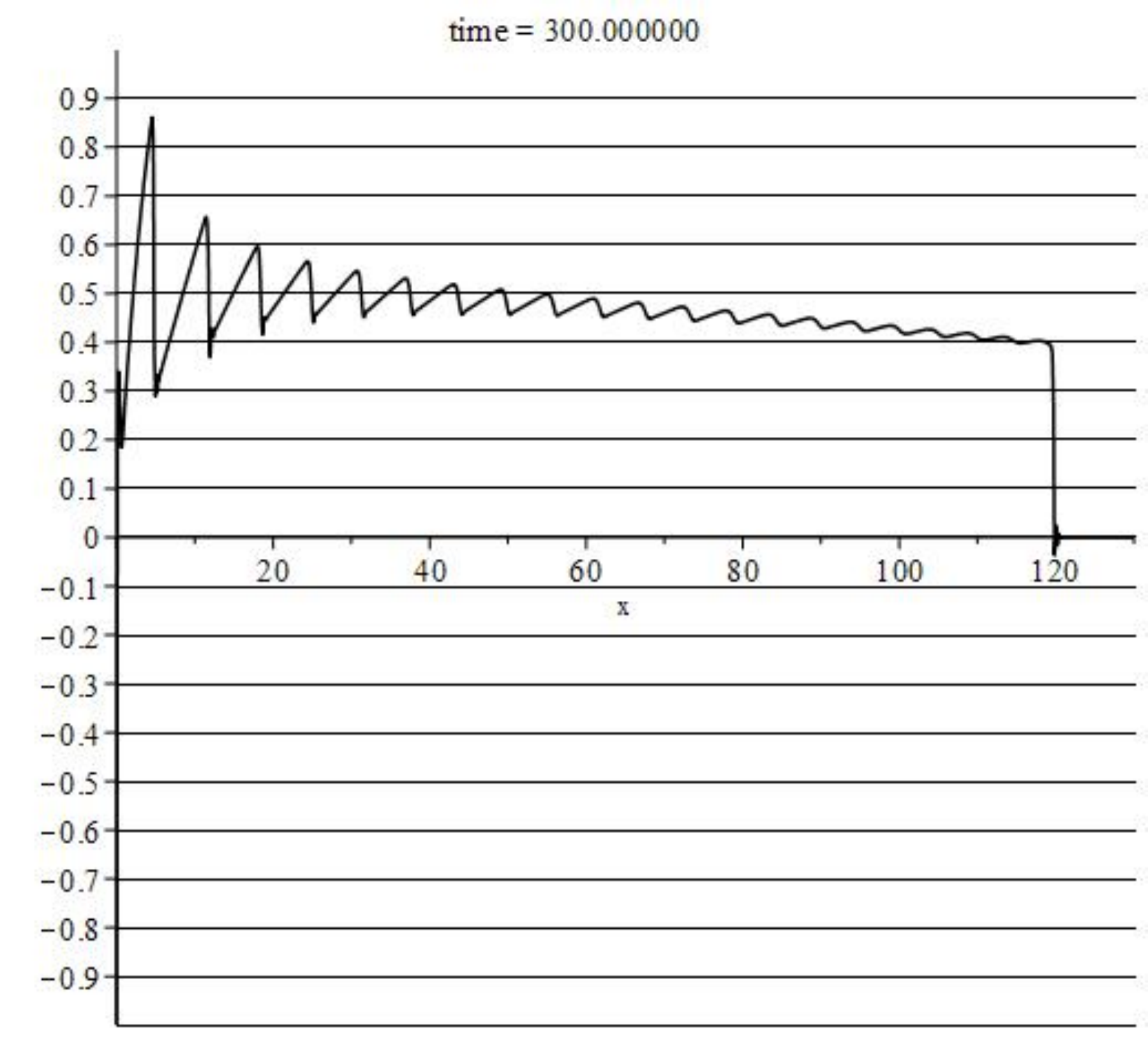}
\end{minipage}
\begin{minipage}{14pc}
\includegraphics[width=14pc]{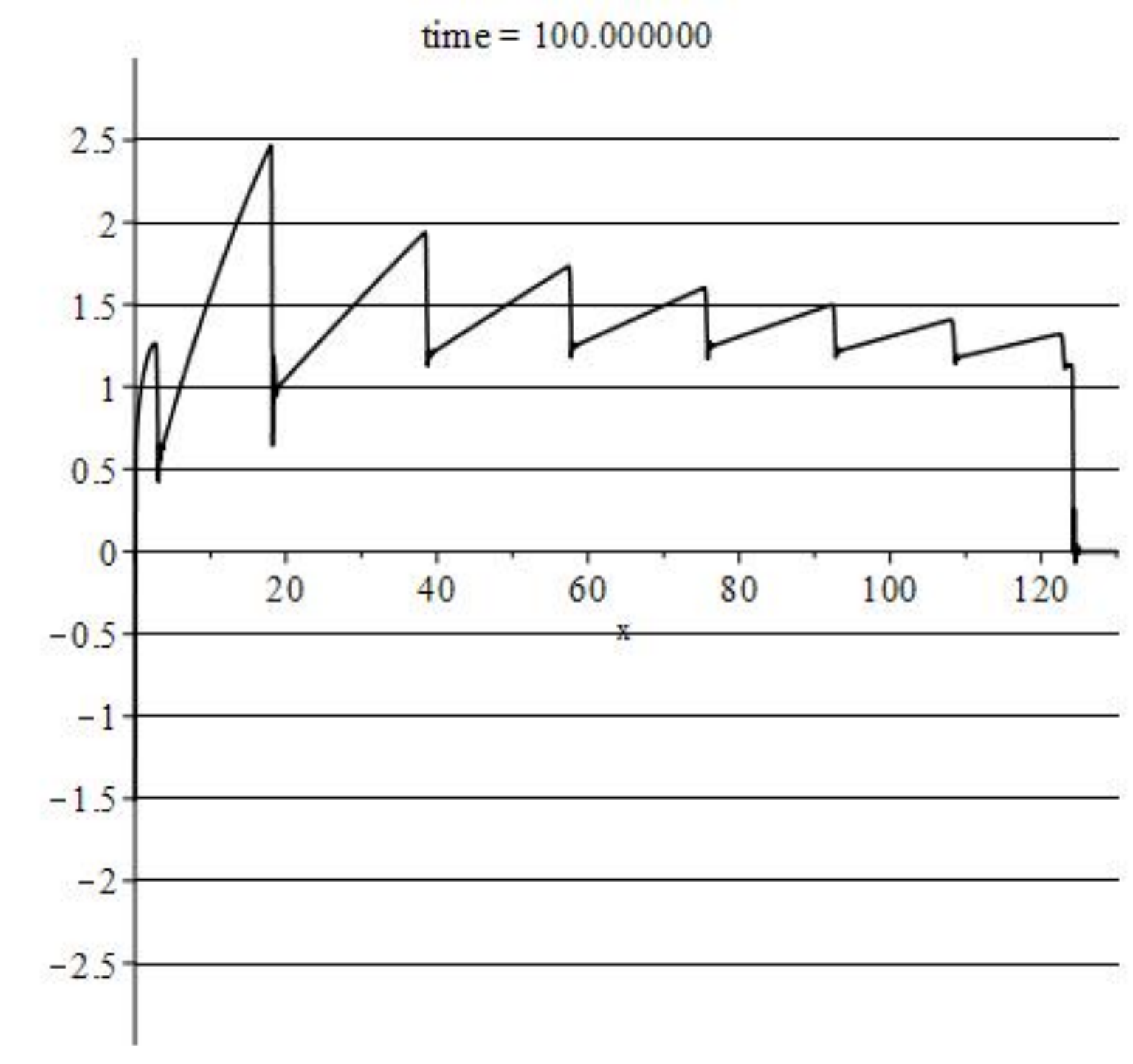}
\end{minipage}
\caption{\textsl{Cylindrical KdV-Burgars, \textbf{Left}: $u_0=\sin t, t=300, \varepsilon=0.1, \delta=0.001$.%\protect\newline
\textbf{Right}:  $u_0=3\sin t, t=100, \varepsilon=0.1, \delta=0.001.$}}
\label{KdV1}
\end{figure}

\begin{figure}[b]
 \begin{minipage}{14pc}
\includegraphics[width=14pc]{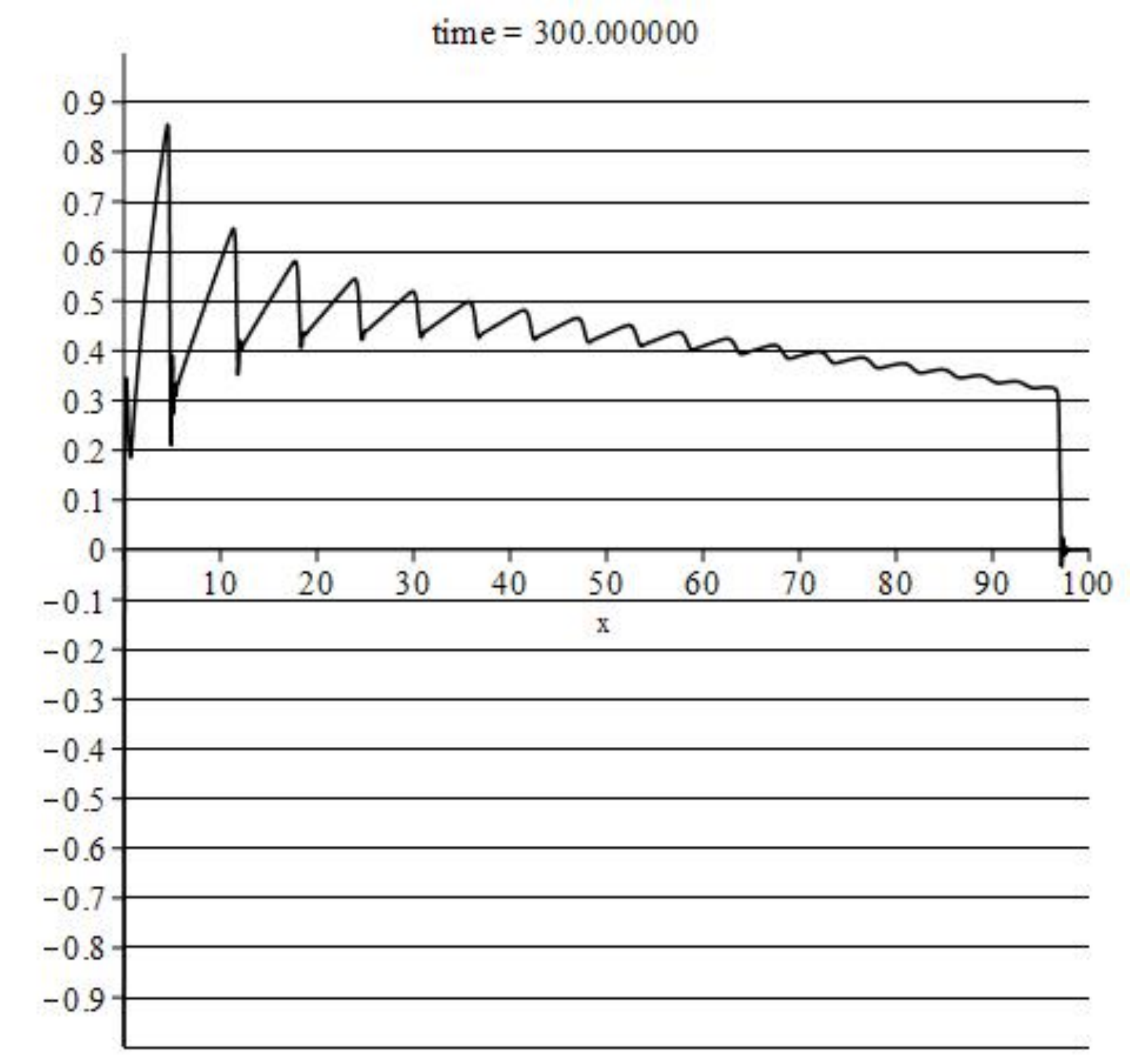}
\end{minipage}
\begin{minipage}{14pc}
\includegraphics[width=14pc]{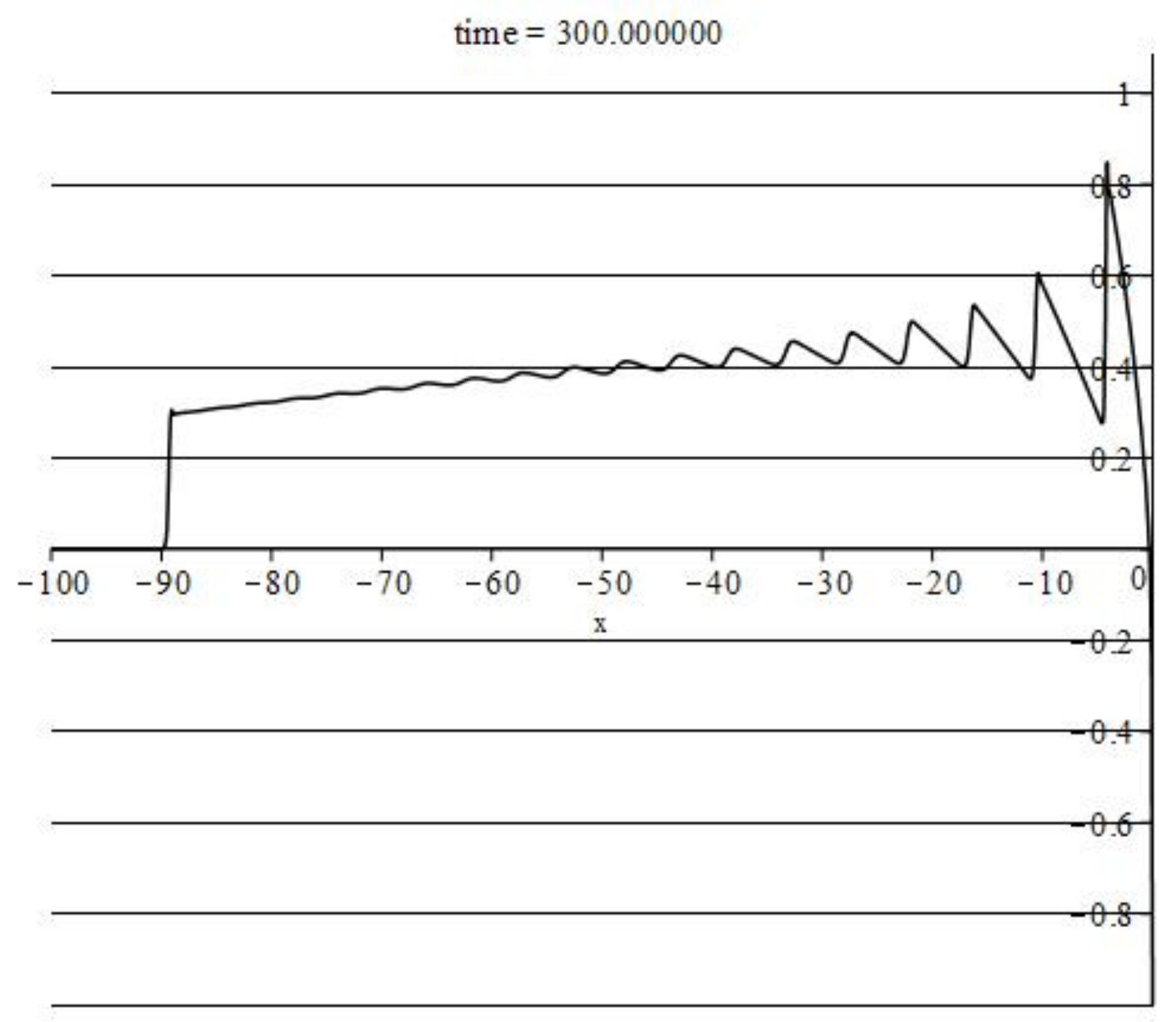}
\end{minipage}
\caption{\textsl{Spherical KdV-Burgers, $u_0=\sin t$, \textbf{Left}: $t=300, \varepsilon=0.1, \delta=0.001.$ %\protect\newline
\textbf{Right}: $u\leftrightarrow -u, t=300, \varepsilon^2=0.02, \delta=0.001$ $\varepsilon^2=0.2$}}
\label{KdV2}
\end{figure}

 Typical graphs for cylindrical and spherical KdV-Burgers, figure \ref{KdV1}, \ref{KdV2}.

\subsection{Overview.} Stronger viscosity effectively damps oscillation and may result in absence of sawtooth effects. Greater frequencies of initial perturbation decay much faster. A signal of a greater amplitude results in increase of velocity and amplitude of travelling signal. After the decay of initial oscillations,  graphs become monotonic declining convex lines, terminating by a shock.

\section{Symmetries and conservation laws}

\subsection{Symmetries}

Since cylindrical and spherical equations explicitly depend on time, their stock of symmetries is scarce.

The algebras of classical symmetries are generated by vector fields:

\begin{eqnarray*}
  X &=& \frac{\partial}{\partial x},\ \\
  Y &=& x\frac{\partial}{\partial x}+2t\frac{\partial}{\partial t}-u\frac{\partial}{\partial u}, \\
  Z &=& \sqrt{t}\frac{\partial}{\partial x}+\frac{1}{4\sqrt{t}}\frac{\partial}{\partial u}, \\
  W &=& \ln(t)\frac{\partial}{\partial x}+\frac{1}{2t}\frac{\partial}{\partial u}.
\end{eqnarray*}
\vspace{3mm}

\begin{tabular}{|l|c|c|}
\hline
Equation& Symmetries & Invariant solutions\\
\hline
&&\\
Cylindrical Burgers & $X, Y, Z$ & $\frac{C}{\sqrt(t)}, \frac{(x+4C)}{4t}$, $x^{-1}f(\frac{t}{x^2}) $\\
&& \small{for some} $f$\\
\hline
&&\\
Cylindrical KdV-Burgers &$X, Z$&$\frac{C}{\sqrt(t)}, \frac{(x+4C)}{4t}$\\
&&\\
\hline
&&\\
Spherical Burgers&$X, Y, W$&$\frac{C}{t},  \frac{x+2C}{2t\ln(t)}$, $x^{-1}f(\frac{t}{x^2}) $\\
&&\small{for some} $f$\\
\hline
&&\\
Spherical KdV-Burgers & $X, W$ & $\frac{C}{t},  \frac{x+2C}{2t\ln(t)}$\\
&&\\
\hline
\end{tabular}

\subsection{Conservation laws}

First rewrite equations \eqref{01} -- \eqref{03} into an appropriate, conservation law form

\begin{equation}\label{04}
    [t^n\cdot u]_t=[t^n\cdot(-u^2+\varepsilon^2 u_{x}+\delta u_{xx})]_x,
    \end{equation}
$n=0,\; 1/2,\; 1$ for flat, cylindrical and spherical cases correspondingly.

Hence for solutions of the above equations we have

\begin{equation}\label{08}
\oint\limits_{\partial\mathcal{ D}} t^n\cdot[u\,dx+(\varepsilon^2 u_x-u^2+\delta u_{xx})\,dt] =0,
\end{equation}

where $\mathcal{D}$ is  a rectangle
\[\{0\leq x\leq L,\;0\leq t\leq T\}.\]

Bearing in mind the initial value/boundary conditions $u(x,0)=u(+\infty,t)=0$, for $L=+\infty$ the integrals read

\[
\int\limits_{+\infty}^0 T^n u(x,T)\,dx+\int\limits_T^0 t^n(\varepsilon^2u_x(0,t)-u^2(0,t)+\delta  u_{xx}(0,t))\,dt=0.
\]

Thus

\begin{equation}\label{10}
\int\limits_0^{+\infty} u(x,T)\,dx=\frac{1}{T^n}\int\limits_0^T t^n(-\varepsilon^2u_x(0,t)+u^2(0,t)-\delta u_{xx}(0,t))\,dt.
\end{equation}
Subsequently

\begin{equation}\label{11}
\frac{1}{T}\int\limits_0^{+\infty} u(x,T)\,dx=\frac{1}{T}\int\limits_0^T \frac{1}{T^n}t^n(-\varepsilon^2u_x(0,t)+u^2(0,t)-\delta  u_{xx}(0,t))\,dt.
\end{equation}
The right-hand side of \eqref{11} is the mean  value right-hand side of \eqref{10}.

It can be computed in some simple cases or estimated.

\subsection{Constant boundary conditions}

\begin{figure}[h]
 \begin{minipage}{14pc}
\includegraphics[width=14pc]{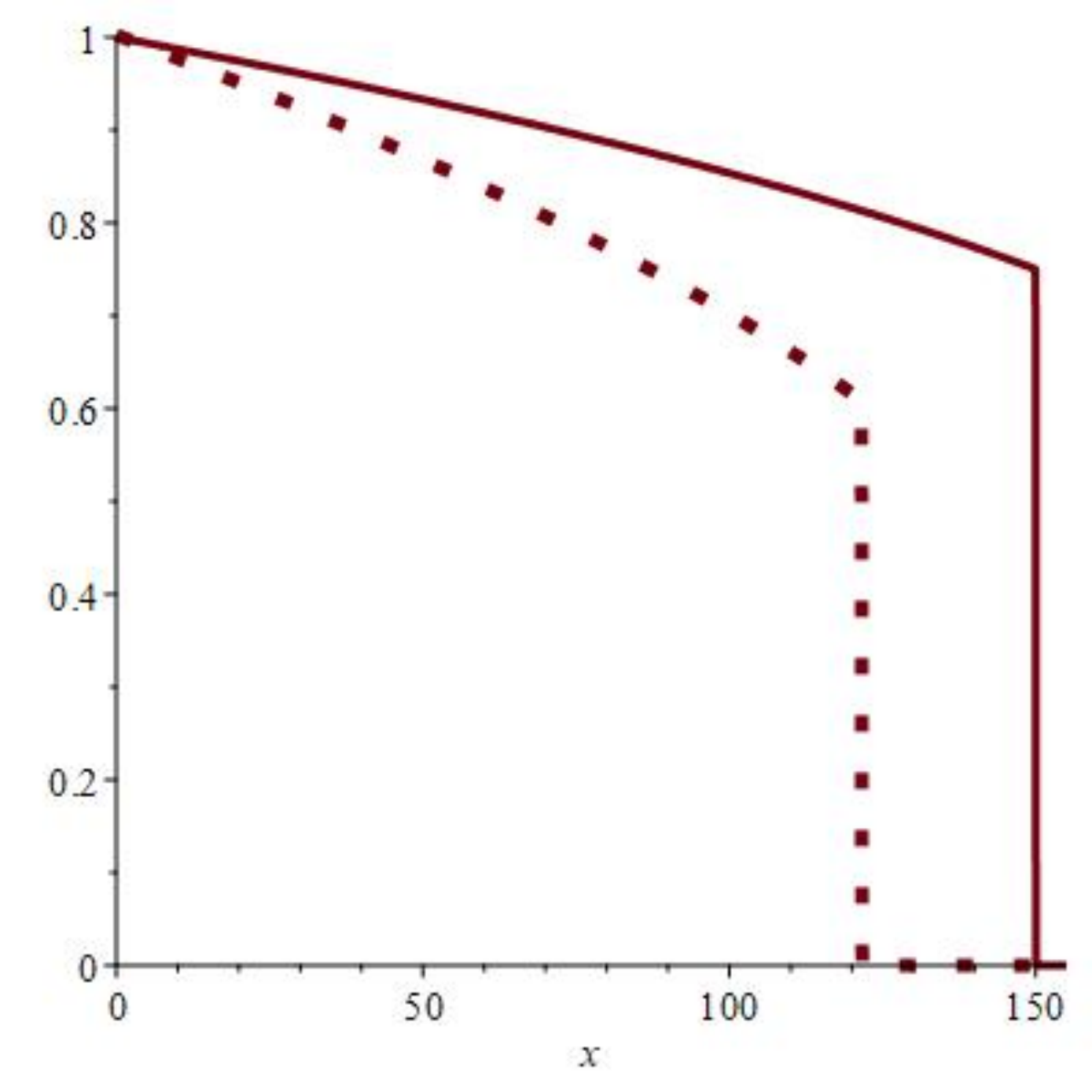}
\end{minipage}
\begin{minipage}{14pc}
\includegraphics[width=14pc]{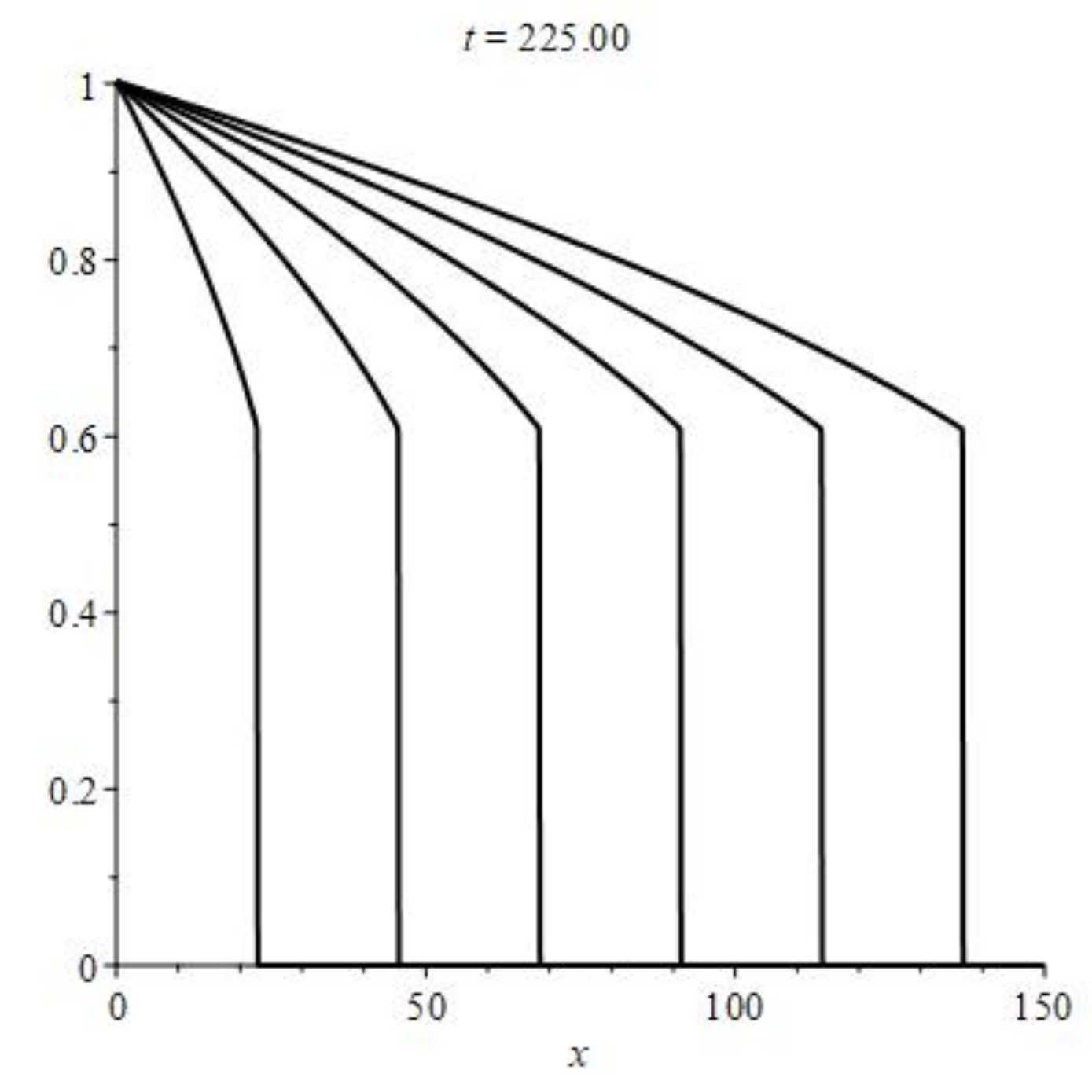}
\end{minipage}
\caption{\textsl{Constant boundary solutions to Burgers equation, $\varepsilon=0.1, t=200. $ \textbf{Left}: Solid line --- cylindrical, dots line --- spherical. %\protect\newline
\textbf{Right}: A trace of movement to the right of the spherical solution at moments $t=37.5\cdot k, k=1\dots 6$}}
\label{const}
\end{figure}

Consider boundary condition $u(0,t)=M$. The graphs of solution are shown on figure \ref{const}, left  (compare the rate of decay caused solely by the spaces dimensions.)

For the resulting compression wave $u_x(0,t)=0$ and the right-hand side of \eqref{11} equals
\begin{equation}\label{ConstRight}
\frac{1}{T}\int\limits_0^T \frac{M^2}{T^n}t^n\,dt=\frac{M^2}{n+1}
\end{equation}

As the figures \ref{B1} --- \ref{KdV2}
show, for periodic boundary condition, after the decay of initial oscillations,  graphs become monotonic convex lines that begin approximately at the hight $A/2$  and broking at $x=V\cdot T$ and at the height $V$.
These monotonic lines are very similar to the graphs or constant-boundary solutions.

\subsection{"Homothetic" solutions}

Looking at the solution's graph animation one can clearly see (eg, on figure \ref{const}, right) that the monotonic part and its head shock develops as a homothetic transformation of the initial configuration.
So we seek solutions of the form $u(x,t)=y(\frac{x}{t})$. Substituting it into equations \eqref{01} -- \eqref{03} we get the equation

\begin{equation}\label{111}
 -y'\frac{x}{t^2}+\frac{ny}{t}=\frac{2yy'}{t}+\frac{\varepsilon^2 y''}{t^2}+\frac{\delta y'''}{t^3},
\end{equation}
or
\begin{equation}\label{112}
 -\xi y'+ny=2yy'+\frac{\varepsilon^2 y''}{t}+\frac{\delta y'''}{t^2},
\end{equation}
for $y=y(\xi)$ and $n=0,\;1/2,\;1$. For $t$ large enough we may omit last two summands.  It follows that appropriate solutions  of the above ordinary differential equations are

\[
u_1(x,t)=C_1,\; C_1\in \mathbb{R},\; n=0,\mbox{ for flat waves equation;}
\]

\[
u_2(x,t)=-\frac{2+\sqrt{C_2\xi+4}}{C_2},\; C_2\in \mathbb{R},\; n=\frac{1}{2},\; \mbox{ for cylindrical and}
\]

\[
u_3(x,t)=\exp\left(\LambertW\left(-\frac{\xi}{2}e^{-\frac{C_3}{2}}\right)+\frac{C_3}{2}\right),\; C_3\in \mathbb{R},\; n=1
\]
for spherical equation.

Let $V$ is the velocity of the signal propagation in the medium. Since at the head  shock $x=Vt$ and $u=V$ we obtain
the condition for finding $C_i$. It is $y(V)=V$. It follows then that $C_1=V,\; C_2=-\frac{3}{V},\; C_3=\ln(V)+\frac{1}{2}$.

For flat waves it corresponds to a travelling wave solution of the classical Burgers equation.

For  the cylindrical waves the monotonic part is given by \[u_2=\frac{1}{3}\left(2V+V\sqrt{4 -\frac{3x}{Vt}}\right);\]
for spherical waves
\[u_3=V\sqrt{e}\exp\left(\LambertW\left(-\frac{x}{2 V t\sqrt{e}}\right)\right).\]

Note that

\begin{equation}\label{connect}
u_2|_{x=0}=\frac{4V}{3}\mbox{ and }u_3|_{x=0}=V\sqrt{e}\approx 1.65 V.
\end{equation}

These formulas show that the velocity is proportional to the amplitude at the start of oscillation. And it does not depend on frequency that together with amplitude define the oscillating part of solutions; more on that below.

\begin{figure}[h]
 \begin{minipage}{12pc}
\includegraphics[width=12pc]{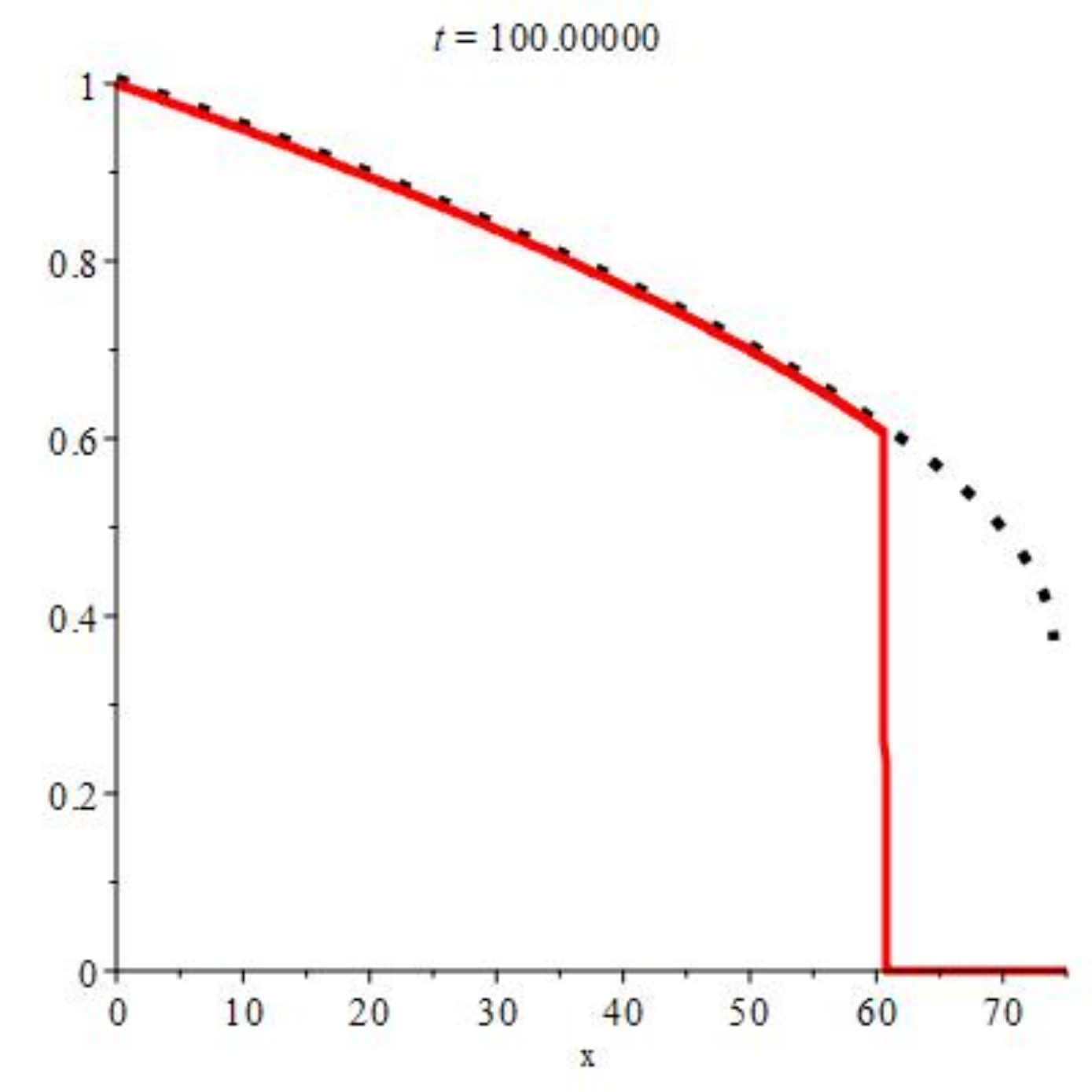}
\end{minipage}
\begin{minipage}{16pc}
\includegraphics[width=16pc]{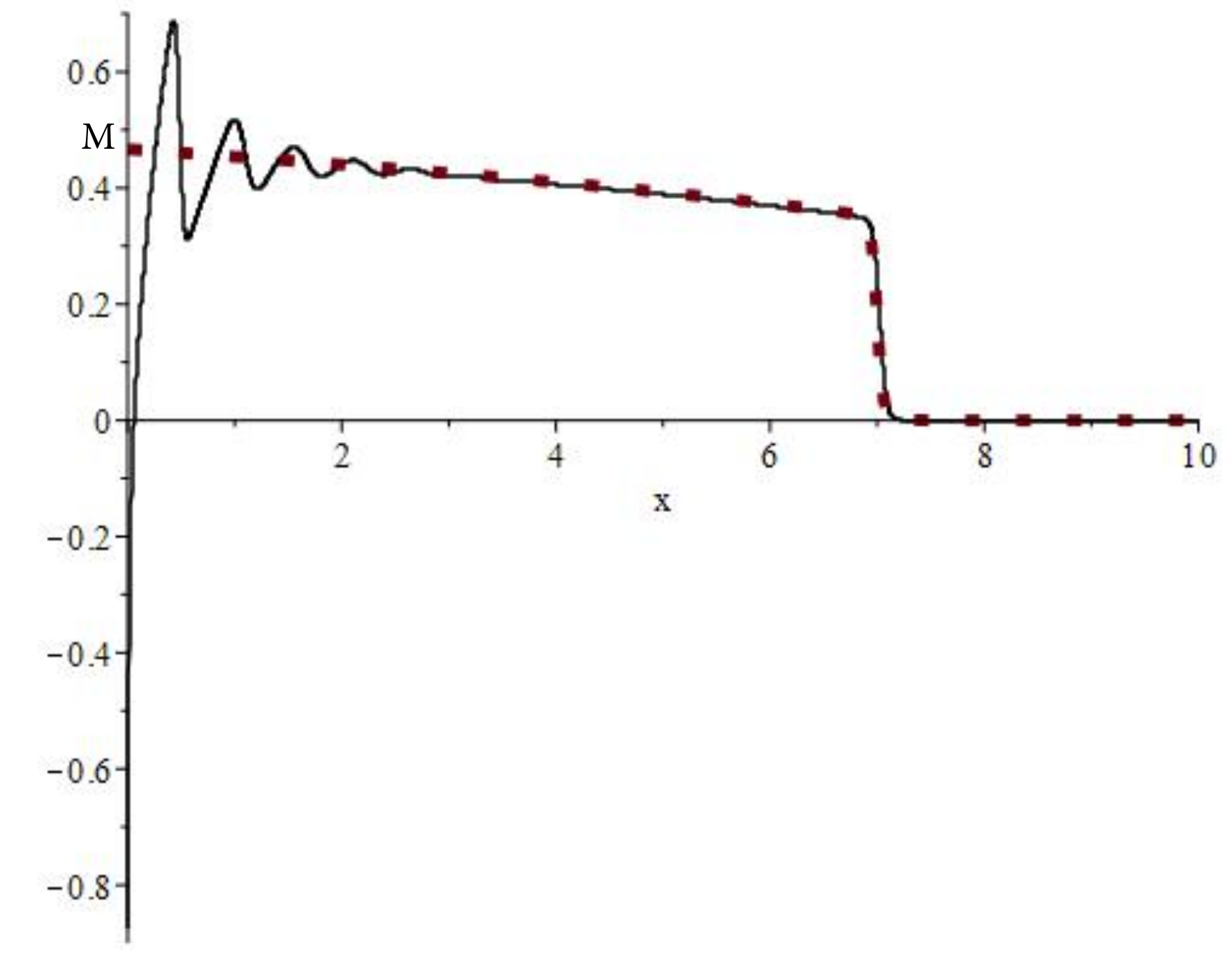}
\end{minipage}
\caption{\textsl{\textbf{Left}: Solid line --- solution to \eqref{t=100}, dots line --- its $u_2$ approximation.    %\protect\newline
\textbf{Right}: Solid line--- solution to \eqref{t=20}, dots line --- its $\tilde u_2$ approximation; both at $t=20$.}}
\label{compare1}
\end{figure}

The corresponding graphs ideally coincide with the graphs obtained by numerical modelling; for instance see comparison  to the solution at ($t=100$) for the problem
\begin{equation}\label{t=100}
 u_t=0.01u_{xx}-2uu_x-u/t,\; u(0,t) = 1, u(75,t) = 0, u(x,0) = 0
\end{equation}
on figure \ref{compare1}, left.

Yet the smooth part of the periodic boundary solution ends with a break, which travels with a constant velocity and amplitude, very much
like a head of the Burgers' travelling wave (TWS) solution \eqref{BTWS}. A rather natural idea is to truncate a homothetic solution, multiplying it by a (normalized) Burgers TWS. Namely, put

\begin{itemize}
  \item For  the cylindrical waves
\begin{equation}\label{Lego1}
\tilde u_2=\frac{1}{2}[1-\tanh(\frac{V}{\varepsilon^2}(x-Vt))]\cdot\frac{1}{3}\left(2V+V\sqrt{4 -\frac{3x}{Vt}}\right);
\end{equation}
  \item for spherical waves
\begin{equation}\label{Lego2}
\tilde u_3=\frac{1}{2}[1-\tanh(\frac{V}{\varepsilon^2}(x-Vt))]\cdot V\sqrt{e}\exp\left(\LambertW\left(-\frac{x}{2 V t\sqrt{e}}\right)\right).
\end{equation}
\end{itemize}

This construction produces an approximation of an astonishing accuracy, see figure \ref{compare1}, right;
this figure corresponds to the cylindrical Burgers problem
 \begin{equation}\label{t=20}
 u_t=0.01u_{xx}-2uu_x-u/2t,\; u(0,t) =\sin 10t , u(10,t) = 0, u(x,0) = 0.
\end{equation}

Moreover, it is evident that the graphs of $\tilde u_2,\, \tilde u_3$ neatly represent the median lines of the approximated solutions on their whole range. Recall that these medians may be evaluated independently via the right-hand side of \eqref{11}.

Now evaluate  the trapezoid area under $\tilde u_2,\, \tilde u_3$ graphs:

\begin{itemize}
  \item For cylindrical equation \[\int_{0}^{Vt}\left[\frac{[1-\tanh(\frac{V}{\varepsilon^2}(x-Vt))]}{2} \frac{1}{3}\left(2V+V\sqrt{4 -\frac{3x}{Vt}}\right) \right]dx=\frac{32}{27}V^2t;\]
  \item for spherical equation
  \begin{multline}
  \int_{0}^{Vt}\left[\frac{[1-\tanh(\frac{V}{\varepsilon^2}(x-Vt))]}{2} V\sqrt{e}\exp\left(\LambertW\left(\frac{-x}{2 V t\sqrt{e}}\right)\right) \right]dx\\
      =\frac{V^2t\cdot e}{2}.
      \end{multline}
\end{itemize}

Hence the mean  value of the left-hand side of \eqref{11} can be estimated as follows. Since the signal from $x=0$ spreads, after decay of oscillations, to the right with  a constant speed $V$ and the same constant amplitude $V$ at the shock,  and it is very well approximated by an appropriate homothetic solution, we get

\begin{equation}\label{curve}
\frac{1}{T}\int\limits_0^{+\infty} u(x,T)\,dx=\frac{1}{T}\int\limits_0^{VT} u(x,T)\,dx
\approx \left\{\begin{array}{l}
\frac{32}{27}V^2 \mbox{ in cylindrical case} ;\\[3pt]
\frac{V^2\cdot e}{2} \mbox{ in spherical case},
\end{array}\right.
\end{equation}
This mean value can be also evaluated numerically. In the case illustrated by figure \ref{B1} the direct numerical evaluation differs from the estimation \eqref{curve} by $1\%$.

For constant-boundary waves, it follows from \eqref{ConstRight} that

\begin{equation}
\frac{M^2}{n+1}=\left\{\begin{array}{l}
\frac{32}{27}V^2 \mbox{ in cylindrical case} ;\\[3pt]
\frac{V^2\cdot e}{2} \mbox{ in spherical case},
\end{array}\right.
\label{linear}
\end{equation}
see \eqref{ConstRight}; of course this result coincides with \eqref{connect}. So the mean value $M$ (see it on figure \ref{compare1}) of arbitrary solution at the start  of oscillations (or in a vicinity of the oscillator) is linearly linked to the velocity of the head shock.

But to find this mean value for an arbitrary border condition is a tricky task, because the integrands $u_x$ and $u_{xx}$  of the right-hand side of \eqref{11} have numerous breaks.

Yet numerical experiments show (eg, see figure \ref{KdV1}), that for the $u|_{x=0}=A\sin(t)$ boundary condition such a value is $M\approx A\cdot a$, where $a\approx 0.467$ is the mean value for $1\cdot\sin(t)$ condition. That is, $M$ depends on $A$ almost linearly.

Note, that this  value may be obtained via the velocity of the head shock, which, in its turn, can be measured with great accuracy by the distance passed by the head shock after a sufficiently long time.

\section*{Conclusion}

We obtained a way to foretell the form of the head shock with a great accuracy: the links in the chain of causations are as follows. First, using the boundary conditions we find, if approximately, the initial mean value of the solution by the formulae \eqref{curve}, \eqref{linear}. This value defines the form of the declining homothetic part of the solution, in particular its velocity and median line. When the amplitude of this declining part reaches the value if velocity, the solution jumps to zero value by the scenario of the Burgers travelling wave and becomes a part of  a homothetic head shock. In vicinity of the boundary oscillations occur around this homothetic line; their longevity, both in time and space, and whether they have a sawtooth form, is defined by the relations between the  amplitude
and frequency of forcing oscillations and viscosity and dispersive characteristics of the media. The exact dependencies are now investigated; results will be published elsewhere.

\subsection*{Acknowledgement}

This work was partially supported by the Russian Basic Research Foundation grant 18-29-10013.

\end{document}